\documentclass[11pt,a4paper,nofootinbib]{article} 
 \pdfoutput=1
\usepackage[utf8]{inputenc}
\usepackage{jheppub}
\usepackage{floatflt,rotate, cancel}
\usepackage{mathrsfs}
\usepackage{amssymb, amsmath, mathtools}
\usepackage[most]{tcolorbox}
\usepackage{color}
\usepackage{graphicx}
\usepackage{subcaption, caption}
\usepackage{multirow}
\usepackage{float}
 \usepackage{cleveref}
 \usepackage{soul}
\usepackage{jheppub}

\title{Generalized momentum operators from Fourier transform correspondence}

\author{Siddharth Dwivedi}

\affiliation{Regional Institute of Education (RIE) Bhubaneswar (NCERT),\\
Bhubaneswar, Odisha -751022
 }

\emailAdd{dwivedi.siddharth@gmail.com}
\abstract{ In this work we take a closer look at the algebraic-operator correspondence between the momentum space and the position space which defines the form of the canonical momentum operator in position space in Quantum Mechanics (QM). Starting from the Fourier transform (FT) relationship, we present a Hermitian generalization of the canonical momentum operator in  position space. 
The action of the generalized operator is found to generate a local flow accompanied by position-dependent rescaling, rather than a global translation. Explicit eigenfunctions are obtained for representative cases and are shown to possess a well-defined limit to the plane-wave solution in QM. As an illustration, the infinite square well problem is solved using the generalized operator, yielding a deformed spectrum that has a smooth limit to the standard QM result.
}

\keywords{Eigenvalues, Operators, Quantum Mechanics, Fourier Transform}
\DeclareUnicodeCharacter{2212}{-}

\begin{document}

\maketitle

%%%%%%%%%%%%%%%%%%%%%%%%%%%%%%%%%%%%%%%%%%%%%%%%%%%%%%%%%%%%%%%%%%%%%%%
\section{Introduction}

In standard QM
% can be interpreted  as an application of Fourier Transform (FT) 
the momentum space and the position space operators are connected via the correspondence \cite{merzbacher1998quantum, Sakurai:2011zz, griffiths_introduction_2018} 
\begin{align}
 ({\rm Momentum ~space})~ \hslash q \leftrightarrow -i \hslash \cfrac{d}{dx} ~({\rm Position~ space}) \label{eq:Sec1_mom_pos_1}
\end{align}
If we consider the (wave)functions in position and momentum space, represented by $\psi(x)$ and $\phi(q)$ respectively, then they are connected by FT via

%\begin{tcolorbox}
\begin{align}
\psi(x) = \cfrac{1}{\sqrt{2 \pi}} \int_{-\infty} ^{\infty} \phi(q) ~e^{iqx} ~dq \label{eq:Sec1_FT}\\
\phi(q) = \cfrac{1}{\sqrt{2 \pi}} \int_{-\infty} ^{\infty} \psi(x) ~e^{-iqx} ~dx \label{eq:Sec1_invFT}
\end{align}
 %\end{tcolorbox}
%%%%%%%%%%%%%%%%%%%%%%%%%%%%
Using Eq.(\ref{eq:Sec1_invFT}), we can derive the
$ q \leftrightarrow -i \dfrac{d}{dx}$ correspondence as follows:
\begin{align}
q ~\phi(q) = \cfrac{1}{\sqrt{2 \pi}} \int_{-\infty} ^{\infty} q ~\psi(x) ~e^{-iqx} ~dx = \cfrac{1}{\sqrt{2 \pi}} \int_{-\infty} ^{\infty} i~  \psi(x) ~\frac{d}{dx}(e^{-iqx}) ~dx,\nonumber
\end{align}
Doing integration by parts we can write,
\begin{align}
 q ~\phi(q) =\cfrac{1}{\sqrt{2 \pi}} \int_{-\infty} ^{\infty} i \left[  ~\frac{d}{dx} \left(\psi(x) e^{-iqx} \right) - \left(~\frac{d \psi(x)}{dx}  \right) e^{-iqx} \right] ~dx\nonumber
\end{align}
Assuming $\psi(x)$ to be a well-behaved function that vanishes as $x \to \pm{\infty}$, the first integral gives zero (vanishing boundary term). Thus we can write
%\begin{tcolorbox}
\begin{align}
 q ~\phi(q) =\cfrac{1}{\sqrt{2 \pi}} \int_{-\infty} ^{\infty} ~ \left(-i ~\frac{d \psi(x)}{dx}  \right) e^{-iqx}  ~dx
 \label{eq:Sec1_mom_pos_2}
\end{align}
%\end{tcolorbox}
Thus we recover the correspondence in Eq.(\ref{eq:Sec1_mom_pos_1}).

In what follows, we take a point of view where we generalize the situation by assuming the relationship between $\phi(q)$ and $\psi(x)$ to be
%\begin{tcolorbox}
\begin{align}
 ~\phi(q) = \cfrac{1}{\sqrt{2 \pi}} \int_{-\infty} ^{\infty} ~\psi(x)~G(q,x) ~dx \label{eq:Sec1_gen_invFT_1}
\end{align}
%\end{tcolorbox}
where $G(q,x) = e^{-iqx}$ if the above definition has to be the familiar FT relation.
%%%%%%%%%%%%%%%%%%%%%%%%%%%%%%%%%%%%%%%%%%%%%%%%%%%%%%%%%%%%%%%%%%%%%%%%%%%%%%%%%%%%%
We now ask the following question: If instead of Eq.(\ref{eq:Sec1_mom_pos_1}) we assume a generalized form for the $q-x$ correspondence to be 
\begin{align}
q \leftrightarrow f(q, x) \frac{d}{dx}, 
\end{align}
then what are the possible solutions for the function $f(q,x)$, consistent with $G(q,x) = e^{-iqx}$? This is derived next in Section \ref{section2:solving_f(q,x)} which leads us to the Hermitian generalization of the momentum operator in position basis. In Section \ref{section3:eignfunction} we solve for the eigenfunctions of the generalized operator. 
%This is followed by a discussion of the generalization of the fundamental commutator in Section \ref{section:fundamental_commutator} and the resulting considerations. 
Subsequently in Section \ref{section:square_well_potential}, as an illustration, we solve for the eigenfunctions and eigenspectrum of the 'generalized' Hamiltonian (using the generalized momentum operator) with the same boundary conditions as for the infinite square well problem in QM. We finally conclude with a summary and discussion of our results in Section \ref{section:summary}. It should be noted that in pursuing this generalization and in what follows, we will think of $q$ as a Fourier domain variable (with the dimension of inverse length) to distinguish it from the eigenvalue variable (denoted by $k$) characterizing the eigenspectrum of the generalized momentum operator.

%In pursuing this generalization and in what follows we will think of $q$ as a Fourier domain variable (with the dimension of inverse length) to distinguish it from the eigenvalue variable $k$ characterizing the eigenspectrum of the generalized momentum operator. It is worth emphasizing that the quantity $\hslash k$ does not represent the kinematic momentum of a physical particle but just characterizes the eigenvalue of the generalized momentum operator. This is due to the fact that the generalized operator is not just a generator of translations but has a different action as we will understand in the following sections.  

%%%%%%%%%%%%%%%%%%%%%%%%%%%%%%%%%%%%%%%%%%%%%%%%%%%%%%%%%%%%%%%%%%%%%%%%%%%%
\section{\boldmath{Solving for $f(q,x)$}}
\label{section2:solving_f(q,x)}
Again, as before, using the generalized relation given in Eq. (\ref{eq:Sec1_gen_invFT_1}) we have,
\begin{align}
 q ~\phi(q) = \cfrac{1}{\sqrt{2 \pi}} \int_{-\infty} ^{\infty} q ~\psi(x) ~G(q,x) ~dx = \cfrac{1}{\sqrt{2 \pi}} \int_{-\infty} ^{\infty} \left(f(q, x) ~\frac{d \psi}{dx}  \right) G(q,x) ~dx
 \label{eq:Sec2_gen_invFT_1}
\end{align}
The last equality comes because we are starting with the assumption that $q \leftrightarrow f(q, x) \cfrac{d}{dx} $. Doing integration by parts on Eq.(\ref{eq:Sec2_gen_invFT_1}) we can write,
\begin{align}
q \phi(q) =\cfrac{1}{\sqrt{2 \pi}} \int_{-\infty} ^{\infty}  \left[ \frac{d}{dx} \left(\psi(x) f(q,x) G(q,x) \right) - \left( \psi(x) \frac{df}{dx} G(q,x) \right) - \left(\psi(x) f(q,x) \frac{d G}{dx}  \right) \right] dx
\label{eq:Sec2_gen_invFT_2}
\end{align}
Here again, the first term gives a boundary term which vanishes for well behaved $\psi(x)$ and we get,
\begin{align}
 q \phi(q) =-\cfrac{1}{\sqrt{2 \pi}} \int_{-\infty} ^{\infty}  \left[ \left( \psi(x) ~\frac{d f}{dx} G(q,x) \right) + \left(\psi(x)~ f(q, x)~ \frac{d G}{dx}  \right) \right] dx
\label{eq:Sec2_gen_invFT_3}
\end{align}
Comparing Eqs. (\ref{eq:Sec2_gen_invFT_1}) and (\ref{eq:Sec2_gen_invFT_3}) and requiring the equality to hold for all well behaved functions $\psi(x)$, we can write
\begin{align}
%  &- \left( \psi(x) ~\frac{d f(q, x)}{dx}~ G(q,x) \right) - \left(\psi(x)~ f(q, x)~ \frac{d  G(q,x)}{dx} \right) = k \psi(x) G(q,x) \nonumber\\
& \frac{d f}{dx}~ G(q,x) + f(q, x)~ \frac{d G}{dx}  = -q G(q,x)
%  &\implies f(q, x)~ \frac{d G(q,x)}{dx}  = -G(q,x) \left(k + \frac{d f(q, x)}{dx} \right) \nonumber\\
%%%%%%%%%%%%%%%%%%%%%%%%%%%%%%%%
%  &\implies \frac{dG(q,x)}{G(q,x)} = -\frac{1}{f(q, x)} \left(k + \frac{d}{dx}f(q, x) \right)
 \label{eq:Sec2_gen_op_func_1}
\end{align}

 %Note that we got Eq.(\ref{eq:Sec2_gen_op_func_1}) by setting the integrands in Eqs. (\ref{eq:Sec2_gen_invFT_1}) and (\ref{eq:Sec2_gen_invFT_3}) equal. This in general is not true because equality of two definite integrals does not necessarily imply equality of integrands. But here we wish to seek such functions $f(q, x)$ and $G(q,x)$ for which the integrands are equal. This is a constraint we impose on our desired solutions.

Integrating Eq.(\ref{eq:Sec2_gen_op_func_1}) we have for $G(q,x)$,
%\begin{tcolorbox}
\begin{align}
  {G(q,x) = {\cal M} ~{\rm exp}\left[- \int{\frac{1}{f(q, x)} \left(q + \frac{d f}{dx} \right) dx} \right]},
 \label{eq:Sec2_gen_op_func_2}
\end{align}
%\end{tcolorbox}
where $\cal M$ is an arbitrary complex constant.

Since we want $G(q,x) = e^{-iqx}$, this implies ${\cal M} = 1$ and 
\begin{align}
 %&\int{\frac{1}{f(q, x)} \left(q + \frac{d f}{dx} \right) dx } = iqx \nonumber \\
  \frac{1}{f(q, x)} \left(q + \frac{d f}{dx} \right) = iq
 \label{eq:Sec2_gen_op_func_3}
\end{align}
Using Eq. (\ref{eq:Sec2_gen_op_func_3}) we have two options to get $G(q,x) = e^{-iqx}$:
\begin{itemize}
 \item {\bf Case 1:} $f(q, x) = -i$. Then we can  directly substitute for $f(q, x)$ in Eq.(\ref{eq:Sec2_gen_op_func_2}) and we get $G(q,x) = e^{-iqx}$.

 \item{\bf Case 2:} $f(q, x) \neq -i$. In this case we need to solve Eq.(\ref{eq:Sec2_gen_op_func_3}). Rearranging,we get
 \begin{align}
%   & k + \frac{d f(q, x) }{dx} = ik f(q, x) \nonumber\\
  & \frac{d f}{dx} = iq(f(q, x) + i) \nonumber \\
%   & \implies \frac{d f(q, x)}{f(q, x) + i} = ik ~dx \nonumber \\
 % & \implies \int{\frac{d f(q, x)}{f(q, x) + i}} = iq \int dx + C_1  \nonumber
 \end{align}
%Here $C_1$ is an arbitrary complex integration constant. 
Integrating, we have for $f(q, x)$,
%\begin{tcolorbox}
\begin{align}
  f(q, x) = {\cal C} e^{iqx} - i,
 \label{eq:Sec2_f(q, x)}
\end{align}
%\end{tcolorbox}
where ${\cal C}$ is an arbitrary complex constant and ${\cal C} \neq 0$. In a more general sense ${\cal C}$ can be a function of the Fourier scale variable $q$ but for simplicity hereafter we consider ${\cal C}$ to be independent of $q$.
\end{itemize}
Thus,  we get the generalized correspondence to be $q \leftrightarrow ({\cal C} e^{iqx} - i) \cfrac{d}{dx}$. Taking cue from this we propose the generalization of the canonical momentum operator in position space to be,
%\begin{tcolorbox}
\begin{align}
  \widehat{p} = \hslash ({\cal C} e^{iqx} - i) \frac{d}{dx}
 \label{eq:Sec2_generalized_p}
\end{align}
%\end{tcolorbox}
The three dimensional generalization, through a similar calculation as above is easy to obtain and comes out to be,
%\begin{tcolorbox}
 \begin{align}
\widehat{\bf p} = \hslash ({\cal C} e^{i\vec{q} \cdot \vec{x}} - i) {\bf \nabla}
 \label{eq:Sec2_generalized_p_3dim}
\end{align}
%\end{tcolorbox}
where $\vec q$ and $\vec x$ are the three dimensional Fourier space vector and position vector respectively.
In the following sections we will only focus on the one dimensional case.
%%%%%%%%%%%%%%%%%%%%%%%%%%%%%%%%%%%%%%%%%%%%%%%%%%%%5

\subsection{Hermitian generalization of the momentum operator}
%%%%%%%%%%%
In general, for the non-vanishing complex constant ${\cal C}$, the generalized momentum operator in Eq.(\ref{eq:Sec2_generalized_p}) is not Hermitian. Its Hermitian conjugate is given by,
%%%%%

\begin{align}
{\widehat {p}}^{\dagger} = - \hslash \left({\cal C}^* e^{-iqx} + i \right)~ \frac{d }{dx} ~ + i \hslash q~ {\cal C}^* e^{-iqx}~
\end{align}
%%%%%%%%%%%%%%%%
 We define the generalized Hermitian operator $\widehat{p}_H$ as
\begin{align}
 & \widehat{p}_H = \frac{1}{2} (\widehat{p} + {\widehat{p}}^{\dagger})  = \frac{\hslash}{2} \left[\left( {\cal C} e^{iqx} - {\cal C}^* e^{-iqx} -2i \right) \frac{d}{dx} + iq {\cal C}^* e^{-iqx} \right]
%   & \implies \widehat{p}_H = \frac{\hslash}{2} \left [\left(2i~ {\rm Im}\left( {\cal C} e^{iqx} \right) -2i \right) \frac{d}{dx} + ik~ {\cal C}^* e^{-iqx} \right] \nonumber
\label{eq:Sec2_general_hermit_op_1}
  \end{align}
  Defining ${\cal C} = a + ib$ with $a,~b$ as real parameters,  we can re-express $\widehat{p}_H$ as,
%%%%%%%%%%%%%%%%%%%%%%%%%%%
 %\begin{tcolorbox}
 \begin{align}
  &\widehat{p}_H = i \hslash \left ( F(q,x) \frac{d}{dx} + G(q,x) \right), \nonumber\\
&  F(q,x) = a \sin(qx) + b \cos(qx) -1, ~~ G(q,x) = \frac{q}{2} (a -ib) e^{-iqx}
 \label{eq:Sec2_general_hermit_op_2}
\end{align}
%\end{tcolorbox}
Here both $a$ and $b$ simultaneously cannot be zero as that would mean ${\cal C} = 0$, which is not a valid assumption for the solution in Eq.(\ref{eq:Sec2_f(q, x)}).
%%%%%%%%%%%%%%%%%%%%%%%%%%%%%%%%%%%%%%%%%%%%%%%%%%%%%%%%%
\subsection{Infinitesimal transformation generated by $\widehat p_H$ }
\label{p_H_as_generator}
In QM the canonical momentum operator $\widehat{p}_0 = -i \hslash \cfrac{d}{dx}$ acts as a generator of translations. Thus, under the action of a unitary operator $\widehat U(d) = e^{\frac{i}{\hslash} d \widehat p_0}$ on a function $\psi(x)$ we get
\begin{align}
\widehat U(d) \psi(x) = \psi(x+ d)
\end{align}
Here $d$ is the translation parameter. Thus the action of the operator generates a global translation for all values of $x$. Considering infinitesimal version of the transformation, we have for the change $\delta \psi(x)$,
\begin{align}
\delta \psi(x) = \psi(x + \epsilon) - \psi(x) = \frac{i}{\hslash} \epsilon \widehat p_0 \psi(x) = \epsilon \frac{d \psi}{dx},
\end{align}
where $\epsilon$ is the infinitesimal translation parameter.
In contrast, the infinitesimal transformation generated by the generalized momentum operator can be given as,
\begin{align}
\delta \psi(x)  = \frac{i}{\hslash} \epsilon \widehat p_H \psi(x) = -\epsilon \left( F(q,x) \frac{d \psi}{dx} + G(q,x) \psi(x) \right)
\label{eq:Sec2.1:p_H_as_generator_1}
\end{align}
This tells us that the action of $\widehat p_H$ does not generate a global translation of the function $\psi(x)$ but instead has two different kinds of effects:
\begin{itemize}
    \item The term $-\epsilon F(q,x) \cfrac{d \psi}{dx}$ corresponds to a local shift via $x \to x - \epsilon F(q,x)$, which implies that different $x$ values undergo different spatial shifts.
    \item The term $-\epsilon G(q,x) \psi(x)$ contributes to a local rescaling of the function via
    \begin{align}
    \psi(x) \to e^{-\epsilon G(q,x)} \psi(x)
    \end{align}
    If we split $G(q,x)$ into real and imaginary parts as $G(q,x) = G_R (q,x) + i G_I (q,x)$, we get
    \begin{align}
    e^{-\epsilon G(q,x)} \psi(x) = e^{-\epsilon G_R(q,x)} e^{-i\epsilon G_I(q,x)} \psi(x),
    \end{align}
    where $e^{-i\epsilon G_I(q,x)} $ contributes to a local phase shift and $e^{-\epsilon G_R(q,x)}$ causes a local amplitude scaling.
\end{itemize}
Thus, in a compact way we can give the infinitesimal action of the generator $\widehat p_H$ as
\begin{align}
\psi(x) \to e^{-\epsilon G(q,x)} \psi(x - \epsilon F(q,x))
\label{eq:Sec2.1:p_H_as_generator_2}
\end{align}
The corresponding action of the generalized operator in the Fourier space is derived in Appendix \ref{sec:Appendix:Mom_op_Fourier_space}.
%%%%%%%%%%%%%%%%%%%%%%%%%%%%%%%%%%%%%%%%%%%%%%%%%%%%%%%%%%%%%%%
\section{Eigenfunctions of $\widehat{p}_H $} \label{section3:eignfunction}
For calculating the eigenfunctions of $\widehat{p}_H$ we consider the following two special cases:
%%%%%%%%%%%%%%%%%%%
\begin{itemize}
%%%%%%%%%%%%%%%%%%%%
 \item  {\bf Case 1: $b = 0$}\\
For this choice, we have
\begin{align}
\widehat{p}_H = i \hslash \left [ \left(a \sin(qx) -1 \right) \frac{d}{dx} + \frac{q a}{2}  e^{-iqx} \right]
\label{eq:Sec3_general_hermit_op_b_0}
\end{align}
The eigenvalue equation is
\begin{align}
& \widehat{p}_H ~\psi_{q, k} (x) = \hslash k ~\psi_{q, k} (x)
\end{align}
where $\psi_{q, k} (x)$ is the eigenfunction corresponding to the operator Fourier parameter $q$ and the eigenvalue $ \hslash k$. 
It is worth emphasizing that the quantity $\hslash k$ does not represent the kinematic momentum of a physical particle but just characterizes the eigenvalue of the generalized momentum operator. This is due to the fact that the generalized operator is not just a generator of translations but has a different action as described in Section \ref{p_H_as_generator}.
Substituting from Eq.(\ref{eq:Sec3_general_hermit_op_b_0}) we have
\begin{align}
 i  \left [ \left(a \sin(qx) -1 \right) \frac{d}{dx} + \frac{q a}{2}  e^{-iqx} \right] \psi_{q, k} (x) = k~ \psi_{q, k} (x)
 \label{eq:Sec3_eigenvalue_eqn_b_0}
\end{align}
%%%%%%%%%%%%%%%%%%%%%%%%%%%%
Integrating Eq.(\ref{eq:Sec3_eigenvalue_eqn_b_0}) we get
\begin{align}
  & \psi_{q, k} (x) = {\cal A}~ {\rm exp}[i k I_1],
  \end{align}
 where $\cal A$ is an arbitrary complex constant and,
 \begin{align}
%I_1 = \int {\cfrac{\left(1 - \cfrac{i a}{2} e^{-iqx}\right)}{(1 - a \sin(qx))} }~ dx
I_1 = \int {\cfrac{1 - \cfrac{i qa}{2k} e^{-iqx}}{ 1- a \sin(qx)} }~ dx
% & {\rm where}~~ I_1 = \int {\cfrac{1}{(1 - a \sin(kx))} }~ dx, ~~~I_2 = \cfrac{-i a}{2} \int \cfrac{ e^{-iqx}}{(1 - a \sin(kx))}~ dx
\end{align}
 For $|a| < 1$ the integrand in $I_1$ does not have a pole for any value of $x$.
Solving for $I_1$ we get for the eigenfunction,
%%%%%%%%%%%%%%%%%%%%%%%%%%%%
%\begin{tcolorbox}
 \begin{align}
 &\psi_{q, k} (x) = \frac{{\cal A}}{\sqrt{1 - a \sin(qx)}} ~{\rm exp} \left[ i \Phi_{q, k}(x) \right], \nonumber \\ 
&  \Phi_{q, k} (x) =   \frac{q x}{2} + \left(\frac{2k}{q} - 1\right)
\frac{1}{ \sqrt{1 - a^2}}  \tan^{-1} \left(\frac{\tan \left(\frac{q x}{2}\right) - a}{\sqrt{1 - a^2}} \right) 
\label{eq:Sec3_p_H_eigenfn_b_0}
\end{align}
%\end{tcolorbox}

The square of the magnitude of the eigenfunction is given by,
%\begin{tcolorbox}
\begin{align}
 | \psi_{q, k} (x) |^2 = \frac{|{\cal A}|^2}{1 - a \sin(qx)}
 \label{eq:Sec3_prob_density_b_0}
\end{align}
%\end{tcolorbox}
We can see that as $a \to 0$, $\psi_{q, k} (x) \to {\cal A}~ e^{i k x}$ and  $ | \psi_{q, k} (x) |^2 \to |{\cal A}|^2$ (constant). For this to agree with the familiar result in QM, we must have ${\cal A} = \cfrac{1}{\sqrt{2 \pi}}$.

%%%%%%%%%%%%%%%%%%%%%%%%%%%%%%%%%%%%%%%%%%%%%%%%

\item  {\bf Case 2: $a = 0$}\\
%%%%%%%%%%%%%%%%%%%%%%%%%%%
In this case the generalized momentum operator is,
\begin{align}
\widehat{p}_H = i \hslash \left [ \left(b \cos(qx) -1 \right) \frac{d}{dx} - \frac{iq b}{2}  e^{-iqx} \right]
\label{eq:Sec3_general_hermit_op_a_0}
\end{align}
The eigenvalue equation becomes
%%%%%%%%%%%%%%%%%%%%%%%%%%%%
\begin{align}
 i  \left [ \left(b \cos(qx) -1 \right) \frac{d}{dx} - \frac{iq b}{2}  e^{-iqx} \right]~ \psi_{q,k} (x) = k~ \psi_{q,k} (x)
 \label{eq:Sec3_eigenvalue_eqn_a_0}
\end{align}
%%%%%%%%%%%%%%%%%%%%%%%%%%%%
Integrating, we have,
\begin{align}
 & \psi_{q, k} (x) = {\cal B}~ {\rm exp}[i k I_2]
 \end{align}
 where ${\cal B}$ is an arbitrary complex constant and,
 \begin{align}
&  I_2 = \int {\cfrac{1 - \cfrac{qb}{2k} e^{-iqx}}{1 - b \cos(qx)} }~ dx
% & {\rm where}~~ I_1 = \int {\cfrac{1}{(1 - b \cos (kx))} }~ dx, ~~~I_2 = \cfrac{-b}{2} \int \cfrac{ e^{-iqx}}{(1 - b \cos(kx))}~ dx
\end{align}
% % % % % % % % % % %
 For $|b| < 1$, integrand in $I_2$ does not have a pole for any value of $x$.
% % % % % % % % % % % % % % % % % %

Solving for $I_2$ we have for the eigenfunction $\psi_{q, k} (x)$,
%%%%%%%%%%%%%%%%%%%%%%
%\begin{tcolorbox}
 \begin{align}
& \psi_{q, k} (x) = \frac{{\cal B}}{\sqrt{1 - b \cos(qx)}} {\rm exp} \left[ i \Gamma_{q, k}(x) \right], \nonumber \\
& \Gamma_{q, k} (x) = \frac{qx}{2} + \left(\frac{2k}{q} - 1\right) 
\frac{1}{ \sqrt{1 - b^2}} \tan^{-1} \left[ \sqrt{\frac{1 + b}{1 -b}}  \tan \left(\frac{q x}{2}\right)  \right] 
\label{eq:Sec3_p_H_eigenfn_a_0}
\end{align}
%\end{tcolorbox}
%%%%%%%%%%%%%%%%%%%%%%%%%
For this case we get,
%\begin{tcolorbox}
\begin{align}
 | \psi_{q, k} (x) |^2 = \frac{|{\cal B}|^2}{1 - b \cos (qx)}
 \label{eq:Sec3_prob_density_a_0}
\end{align}
%\end{tcolorbox}
As before, ${\cal B} = \cfrac{1}{\sqrt{2 \pi}}$ for the results to agree with the familiar result in QM in the limit $b \to 0$.
%%%%%%%%%%%%%%%%
\item {\bf Case 3: The general case, $a \neq 0,~ b \neq 0$}\\
This involves more detailed algebraic manipulations and is not specially informative apart from the fact that the eigenfunction  has a smooth limit to the QM result for $(a, b) \to (0,0)$ while $a^2 + b^2 < 1$. So we do not present here an explicit calculation for the same.
%%%%%%%%%%
\end{itemize}

%%%%%%%%%%%%%%%%%%%%%%%%%%%%%%%%%%%%%%%%%%%%%%%%%%%%%%%%%%%%%%%
\section{A simple test case: The Infinite Square Well Potential }
\label{section:square_well_potential}
As an example to illustrate the modifications brought to the eigenfunctions and eigenvalues, we solve the eigenvalue equation  for the operator $\widehat p^2_H/2m$ with the boundary conditions of an infinite square well potential of width $L$ with potential $V(x)$ such that,
\begin{align}
 & V(x) = 0 ~~~ (0 \leq x \leq L) \nonumber \\
 & V(x) = \infty ~~~ (x < 0 ~~\text{or}~~x > L) 
\end{align}
In the QM limit,  $m$ can be identified with the mass of a particle confined in the well.
Focusing on the case for which $b = 0$, the eigenvalue equation can be given as,
\begin{align}
 {\widehat{p}_H} ^2 ~\psi(x) = 2 m E~ \psi(x)
\end{align}
where $E$ is the allowed eigenvalue. We take $E > 0$ for normalizable solutions in the QM limit ($a \to 0$). Defining $k_0 = \cfrac{\sqrt{2 m E}}{\hslash}$ ($k_0 > 0$), we can trivially factorize the problem as
\begin{align}
 \left(\widehat{p}_H \pm k_0 \right) \left(\widehat{p}_H \mp k_0 \right) \psi(x) = 0
\end{align}
Thus, we have two independent solutions corresponding to
\begin{align}
 \widehat{p}_H ~\psi(x) = \pm k_0 ~ \psi(x)
\end{align}
The two independent solutions are the two eigenfunctions of $\widehat{p}_H$ , $\psi_{q, \pm k_0} (x)$ corresponding to $k = \pm k_0$, {\it i.e.}
\begin{align}
\psi_{q, \pm k_0} (x) =  \frac{{ 1}}{\sqrt{1 - a \sin(qx)}} ~{\rm exp} \left[ i \left(  \frac{q x}{2} + \left(\frac{\pm 2k_0}{q} - 1\right)
\frac{1}{ \sqrt{1 - a^2}}  \tan^{-1} \left(\frac{\tan \left(\frac{q x}{2}\right) - a}{\sqrt{1 - a^2}} \right) \right) \right]
\end{align}

% The $'+'$ sign corresponds to $+ k_0$ and the $'-'$ sign to $-k_0$ respectively. Using the momentum eigenfunction $\psi_k (x)$ derived in Section \ref{sec:eignfn}
We can write the general solution as
\begin{align}
 \psi(x) = {\cal P} ~\psi_{q, +k_0} (x) + {\cal Q}~ \psi_{q, -k_0} (x)
 \label{eq:Sec5_Square_well_eigenfunction_1}
\end{align}
where ${\cal P}$ and ${\cal Q}$ are arbitrary constants to be determined by the boundary conditions of the problem, {\it i.e.} $\psi(x =0) = 0 = \psi(x=L)$.
For $\psi(x =0) = 0$ we have
% \begin{align}
% & ({\cal M} + {\cal N})~{\rm exp} \left[ -i \left(
% \frac{1}{ \sqrt{1 - a^2}}  \tan^{-1} \left(\frac{a}{\sqrt{1 - a^2}} \right) \right) \right] = 0
% \end{align}
% which gives
\begin{align}
& {\cal P} ~e^{-i \frac{2 k_0 \theta_1}{q}} + {\cal Q} ~e^{i \frac{2 k_0 \theta_1}{q}} = 0, ~{\rm where} \nonumber \\
 &\theta_1 = \frac{1}{  \sqrt{1 - a^2}}  \tan^{-1} \left(\frac{ a}{\sqrt{1 - a^2}} \right)
 \label{eq:Sec5_square_well_bc_1}
\end{align}
The other boundary condition $\psi(x=L) = 0$ leads to
\begin{align}
& {\cal P}~ e^{i \frac{2 k_0 \theta_2(L)}{q}} + {\cal Q}~ e^{-i \frac{2 k_0 \theta_2(L)}{q}} = 0, ~{\rm where} \nonumber \\
& \theta_2(x) = \frac{1}{  \sqrt{1 - a^2}}  \tan^{-1} \left(\frac{ \tan \left( \frac{qx}{2} \right)  - a}{\sqrt{1 - a^2}} \right)
 \label{eq:Sec5_square_well_bc_2}
\end{align}
Using Eq.(\ref{eq:Sec5_square_well_bc_1}) and Eq. (\ref{eq:Sec5_square_well_bc_2}) gives,
\begin{align}
& e^{i\frac{4 k_0}{q}(\theta_1 + \theta_2(L))} = 1, ~ {\rm or} \nonumber \\
&\cos \left(\frac{4 k_0}{q}(\theta_1 + \theta_2(L)) \right) =1, ~~ \sin \left(\frac{4 k_0}{q}(\theta_1 + \theta_2(L)) \right) =0
\label{eq:Sec5_square_well_bc_3}
\end{align}
which leads to 
\begin{align}
&\frac{2 k_0}{q} (\theta_1 + \theta_2(L)) = n\pi ~~~~(n =1, ~2, ~3..) \nonumber \\
& \implies k_0 = \frac{ n\pi q \sqrt{1 - a^2} }{2 \tan^{-1} \left( \frac{ \tan \left( \frac{q L}{2} \right) \sqrt{1-a^2}}{1 - a \tan \left( \frac{q L}{2} \right)} \right)}
\label{eq:Sec5_square_well_quantization_1}
\end{align}
Using the definition of $k_0$ in terms of $E$ in Eq. (\ref{eq:Sec5_square_well_quantization_1}) we have,
\begin{align}
E = \frac{n^2 \pi^2 \hslash^2}{8 m} \frac{q^2 (1 - a^2)} { \left[ \tan^{-1} \left( \frac{ \tan \left( \frac{q L}{2} \right) \sqrt{1-a^2}}{1 - a \tan \left( \frac{q L}{2} \right)} \right) \right]^2}
\end{align}
Using Eqns. (\ref{eq:Sec5_Square_well_eigenfunction_1}- \ref{eq:Sec5_square_well_bc_2}) We can rewrite the eigenfunction as
\begin{align}
\psi(x) =  \frac{\cal P}{\sqrt{1 - a \sin(qx)}} ~e^{ i \left(  \frac{q x}{2} - \theta_2(x)\right)} \left(e^{i \frac{2 k_0 \theta_2(x)}{q}} - e^{-i \frac{2 k_0}{q} (2\theta_1 + \theta_2(x)) } \right)
\label{eq:Sec5_Square_well_eigenfunction_2}
\end{align}
Looking at Eq. (\ref{eq:Sec5_Square_well_eigenfunction_2}) we can see that for $n =0$, $k_0 = 0$ and $\psi(x)$ vanishes identically in the domain $0\leq x \leq L$, which does not give a meaningful probabilistic interpretation in the QM limit. Moreover, negative non-zero integer values of $n$ lead to a phase shift of the solution by
\begin{align}
\psi(x) \xrightarrow{n \to -n} -e^{i \frac{4 k_0 \theta_1}{q}} \psi(x),
\end{align}
%where $k_0 > 0$ as given in Eq.(\ref{eq:Sec5_square_well_quantization_1}). 
which does not have any physical consequence for the results in the QM limit. This is why we consider only positive integer values of $n$ in quantizing $k_0$.
In the limit $a \to 0$, we have $\theta_1 \to 0$, $\theta_2(x) \to \frac{qx}{2}$, $k_0 \to \frac{n \pi}{L}$ and the wavefunction $\psi (x) \to 2i {\cal P} \sin \left(\frac{n \pi x}{L} \right)$. For this to match the standard  result in QM \cite{merzbacher1998quantum, griffiths_introduction_2018, Sakurai:2011zz},  we must have 
\begin{align}
 2 i {\cal P} = \sqrt{\frac{2}{L}} ~~{\rm or}~~ {\cal P} = -\frac{i}{\sqrt{2 L}}
\end{align}
%%%%%%%%%%%%%%%%%%%%%%%%%%%%%%%%%%%%%%%%%%%%%%%%%%%%%%%%%%%%%%%
%%%%%%%%%%%%%%%%%%%%%%%%%%%%%%%%%%%%%%%%%%%%%%%%%%%%%%%%%%%

\section{Summary and Conclusions} \label{section:summary}

Starting from the FT relationship between position and momentum space representations in standard QM, we have re-examined the algebraic correspondence that underlies the form of the momentum operator in position space. We derived a generalized correspondence between multiplication in Fourier space and differentiation in position space
%, while retaining the standard Fourier kernel, 
which gave us a continuous family of generalized momentum operators characterized by a Fourier scale ($q$) and two real deformation parameters ($a,~b$).

A Hermitian generalization of the above mentioned operator was constructed and it was found that unlike the canonical momentum operator, the generalized operator does not generate global translations of the function on which it acts, but instead induces a position dependent translation along with a local rescaling. The corresponding eigenfunctions were obtained for representative cases and were shown to possess smooth plane-wave limits as the deformation parameters are taken to zero.

As an illustration of the formalism, the infinite square well problem was solved using the generalized operator, yielding a deformed energy spectrum that continuously connects to the standard QM result. In this sense,
%the canonical momentum operator emerges as a special point within a broader algebraic family characterized by the Fourier scale parameter and deformation parameters.
the canonical momentum operator corresponds to the point $(a=0,b=0)$ within the allowed domain of deformation with a smooth QM limit, {\it i.e.} the open unit disk $a^2 + b^2 < 1$. In other words, in the two dimensional deformation space comprising $a,~b$, QM sits at the center $(0,0)$ and the generalized operators correspond to all other points $(a, b)$ of the open unit disk $a^2 + b^2 < 1$.

\section{Acknowledgment}
We would like to thank Sayan Kar, Biswarup Mukhopadhyaya and Vikash Pandey for their useful comments and insights.

%%%%%%%%%%%%%%%%%%%%%%%%%%%%%%%%%%%%%%%%%%%%%%%%%%%%%%%%%%%%%%%%%%%%%%%%%%%%%%%%%%%%%%%
%\newpage
\appendix
\section{Action of generalized momentum operator in Fourier space}
\label{sec:Appendix:Mom_op_Fourier_space}
Denoting the Fourier space variable by $q'$, the action of the generalized momentum operator on the Fourier domain is defined by
\begin{align}
(\widehat{p}_H \phi)(q') = \cfrac{1}{\sqrt{2 \pi}} \int_{-\infty} ^{\infty} (\widehat{p}_H \psi)(x) ~e^{-iq'x} ~dx 
\label{eq:Appendix_pH_in_q_space_1}
\end{align}
Using the definition of $\widehat{p}_H $ in Eq. (\ref{eq:Sec2_general_hermit_op_2}) we get
\begin{align}
(\widehat{p}_H \phi)(q') = \cfrac{1}{\sqrt{2 \pi}} \int_{-\infty} ^{\infty} \left( F \psi' + G \psi \right) ~e^{-iq'x} ~dx 
\label{eq:Appendix_pH_in_q_space_2}
\end{align}
where $\psi' = \cfrac{d\psi}{dx}$. From Eq. (\ref{eq:Sec1_FT}) we have for $\psi'$,
\begin{align}
\psi'(x) =  \cfrac{1}{\sqrt{2 \pi}} \int_{-\infty} ^{\infty} iq'' \phi(q'')~e^{iq''x} ~dq''
\end{align}
where $q''$ is a dummy Fourier variable. Substituting the above for $\psi'$ in Eq. (\ref{eq:Appendix_pH_in_q_space_2}) and using the integral representation of Delta function, {\it i.e.} $\delta(q) =  \frac{1}{2 \pi} \int_{-\infty} ^{\infty} e^{iqx} dx$, we do the $x$ integration to obtain,
\begin{align}
(\widehat{p}_H \phi)(q') &= i \hslash \Biggl[ \frac{(b-ia)}{2} \int_{-\infty} ^{\infty} ( iq'') \phi(q'') \delta(q''+ q -q') dq'' \nonumber \\
& + \frac{(b+ia)}{2} \int_{-\infty} ^{\infty} ( iq'') \phi(q'') \delta(q''- q -q') dq'' \nonumber\\
& - \int_{-\infty} ^{\infty} ( iq'') \phi(q'')  \delta(q'' -q') dq'' + \frac{q(a-ib)}{2} \int_{-\infty} ^{\infty} \phi(q'') \delta(q''-q' -q) dq''\Biggr] \nonumber \\
\implies (\widehat{p}_H \phi)(q') &= \frac{\hslash}{2} \left[ i(q'-q){\cal C} \phi(q'-q)
-iq' {\cal C^*} \phi(q' + q) + 2 q' \phi(q') \right]
\end{align}
Thus we see that the action of the generalized operator on the Fourier domain leads to mode mixing, and we get the familiar QM result only in the limit ${\cal C} \to 0$, for which $(\widehat{p}_H \phi)(q') =\hslash q' \phi(q')$.
%%%%%%%%%%%%%%%%%%%%%%%%%%%%%%%%%%%%%%%%%%%%%%%%%%%%%%%%%%%

\bibliographystyle{jhep}
\bibliography{QM}
\end{document}